# Enhancing the OPEN Process Framework with Service-Oriented Method Fragments

Mahdi Fahmideh[1], Mohsen Sharifi [2] and Pooyan Jamshidi [3]

[1] Azad University of Science and Research Branch, Tehran, Iran
[2] School of Computer Engineering, Iran University of Science and Technology, Tehran, Iran
m.fahmideh@comp.iust.ac.ir, msharifi@iust.ac.ir
[3] Lero - The Irish Software Engineering Research Centre,
School of Computing, Dublin City University, Dublin, Ireland
pooyan.jamshidi@computing.dcu.ie

**Abstract**: Service-orientation is a promising paradigm that enables the engineering of large-scale distributed software systems using rigorous software development processes. The existing problem is that every service-oriented software development project often requires a customized development process that provides specific service-oriented software engineering tasks in support of requirements unique to that project. To resolve this problem and allow situational method engineering, we have defined a set of method fragments in support of the engineering of the project-specific service-oriented software development processes. We have derived the proposed method fragments from the recurring features of eleven prominent service-oriented software development methodologies using a systematic mining approach. We have added these new fragments to the repository of OPEN Process Framework to make them available to software engineers as reusable fragments using this well-known method repository.

*Keyword: Service-Oriented Software Development, OPEN Process Framework, OPF Repository, Method Fragment, Situational Method Engineering*

# I. INTRODUCTION

Software engineers are currently faced with increasing demands for the development of software systems that are heterogeneous, geographically distributed and dynamic in nature in the sense that system components can be dynamically detached, added or reconfigured at runtime [1]. Service-oriented paradigm has provided the basic concepts and means for development of such software systems. Services as fundamental elements of service-oriented systems play a pivotal role in service-oriented software development. They are self-contained, loosely coupled, platform independent, stand-alone, and autonomous elements that form the underpinning of service-oriented systems [2]. A number of available published services can be composed together to form a large software system. Services collaborate via standard message protocols in a loosely coupled distributed heterogeneous environment. It is thus possible for software engineers to develop service-oriented software systems via composition of discovered services during software construction or execution rather than crudely following traditional phases of analysis, design, and implementation. To take advantage of existing services, service-oriented software developers must perform extra tasks compared to traditional software developers that are specific to service-orientation. Furthermore, software requirements are less known to service-oriented software developers while traditional software developers have more knowledge about software requirements at earlier stages of software development and know the tasks they must perform to satisfy these requirements earlier [3].

Service-oriented *Software Development Methodologies* (*SDM*s) have tried to identify tasks that service-oriented software developers must carry out in addition to tasks carried out in traditional software development methodologies. These extra tasks are specific to *Service-Oriented Software Development* (*SOSD*). Although SDMs have some common features (e.g. cover the same life-cycle phases), they have been proposed for different purposes, ranging from project management to system modernization, and from business analysis to development of technical solutions [4]. Given the variety of existing SDMs, it is hard for software engineers to decide which SDM fits best the specific needs of a project. Furthermore, specific SOSD tasks in service-oriented SDMs are tightly interwoven with traditional tasks making it very hard for developers to extract and assemble the required SOSD tasks in support of requirements of a specific project. This asserts the evidence that there is no universal software development process[1] that is appropriate for all situations [5,6,7,8]. Some of the issues that developers must consider for every situation include organizational maturity and culture, people skills, commercial and development strategies, business constraints, and tools [9,10]. They must therefore construct their own project-specific SDM or software process for the development of their software.

One of the well-known approaches for tailoring SDMs is *Situational Method Engineering (SME)*, wherein a project-specific SDM is constructed from *Reusable Method Fragments* [11,12] or *Method Chunks* [7,13]. To allow the construction of a wide range of project-specific SDMs by developers and method engineers, a repository of method chunks is necessary [5]. An established approach in line with the ideas of SME is the *Object-oriented Process, Environment, and Notation* (*OPEN*) [14,15]. OPEN has a repository of reusable method fragments called *OPF* from which method engineers can select method fragments using suitable construction guidelines. They can then assemble their selected fragments to construct a wide spectrum of project-specific SDMs based on the unique set of requirements of SDMs. Existing method fragments in OPF can be used in the construction of many types of situational SDMs except for service-oriented SDMs. In other words, one of the main shortcomings of OPEN is its lack of support for SOSD. Existing method fragments in OPF repository are mainly intended for *Object-Oriented* (*OO*) software development while method fragments in support of agility and aspect orientations are also forthcoming [36, 37, 38, 39]. Although there are many commonalities between OO software development and SOSD, they have many differences too requiring new method fragments in support of SOSD.

Motivated to enhance OPF repository, we propose a new set of method fragments in this paper in support of SOSD in conformance with the underpinning metamodel standard of OPEN [27] using our previous systematic approach [16]. We have designed these method fragments in such a way to facilitate the engineering of service-oriented SDMs based on OPEN. To do so, we studied the SOSD literature, specifically the development processes of most well-known existing service-oriented SDMs, extracted their recurrent tasks, and presented extracted tasks in the form of method fragments. OPEN CASE tools [17] that manage the OPF repository can import the proposed method fragments as extensions to their existing OPF repository and use them to construct project-specific service-oriented SDM.

---

[1] In this paper, we consider the terms *method*, *methodology, software development methodology, and software development process,* as synonyms.

Having delineated the outline of our research, we have organized the rest of paper as follows. Section II presents the basic concepts underlying our research. Section III presents a brief review of prominent service-oriented SDMs that have been selected as main sources to define new method fragments. Section V explains the way in which appropriate method fragments have been constructed. Section VI presents our proposed method fragments. Section VII identifies the position of these method fragments in the OPEN process framework. Section VIII presents a discussion on the proposed method fragments. Section IX shows the applicability of the new method fragments through presentation of a partial case study. Section X concludes the paper and presents further extensions to the reported research.

## II. BACKGROUND

In this section we briefly review the main concepts underlying our proposition in this paper.

### A. Situational Method Engineering

The prevalent belief that no single software development process can be applicable to all situations is the main reason for the emergence of *Method Engineering* (*ME*). ME was first introduced by Kumar [5] as a software engineering discipline aimed at constructing a project-specific software development process to meet given organizational characteristics and project situations. Brinkkemper [6] elaborated ME definition later to: "The engineering discipline to design, construct, and adapt methods, techniques and tools for the development of information systems". The most well-known subset of ME, namely SME, is concerned with the construction, adaptation or enhancement of a suitable SDM for the project at hand instead of looking for a universal or widely applicable one [5,6,7,8]. In the SME approach, an SDM is constructed from a number of encapsulated and fragmentized methods stored in a repository. Typically, a method engineer goes through the following SME steps to construct a project-specific SDM [19]:
1. Elicitation and specification of specific requirements of target SDM.
2. Selection of a number of most relevant method fragments from the repository based on a number of situational factors highly specific to the particular software development organization and particular situation of the project.
3. Assembly of the chosen method fragments to form a coherent project-specific SDM.

Method engineers can use *Computer Aided Method Engineering* (*CAME*) tools to do the above four steps for saving, restoring, selecting and assembling method fragments [21]. One instance of the SME approach that is highly compatible with the above steps and is extensively used in the development of a wide range of software project types, especially in the OO context, is the OPEN Process Framework [14,15]. Industrial use of OPEN demonstrates its viability in software development [22] so much so that we have been motivated to base our research on OPEN. In the next subsection, we present OPEN in more depth.

### B. OPEN Process Framework as a Foundation for SME

OPEN is the oldest established software development process introduced in 1996 as a result of the integration of three second-generation OO software development SDMs, namely MOSES [232023], SOMA [24] and Firesmith. OPEN is known as one of the most popular software development processes with support for full lifecycle. OPEN has been updated recently to be conformant with ISO/IEC 24744 [25], which is mainly intended for using in the development of software systems or in the construction of project-specific SDMs based on projects' circumstances. A not-for-profit consortium comprising of an international group of methodologists, academics, and CASE tool vendors maintains OPEN [26]. OPEN contains an underpinning metamodel (a model for describing method fragments or software development processes), a rich repository of method fragments, and several kinds of usage guidelines that explain how method engineers can use method fragments. To construct a project-specific SDM, a method engineer selects his/her required method fragments from the OPF repository wherein each method fragment is an instance of OPEN metamodel (Fig. 1). Given our objective in this paper, we study the metamodel of OPEN and its OPF repository in more detail here.

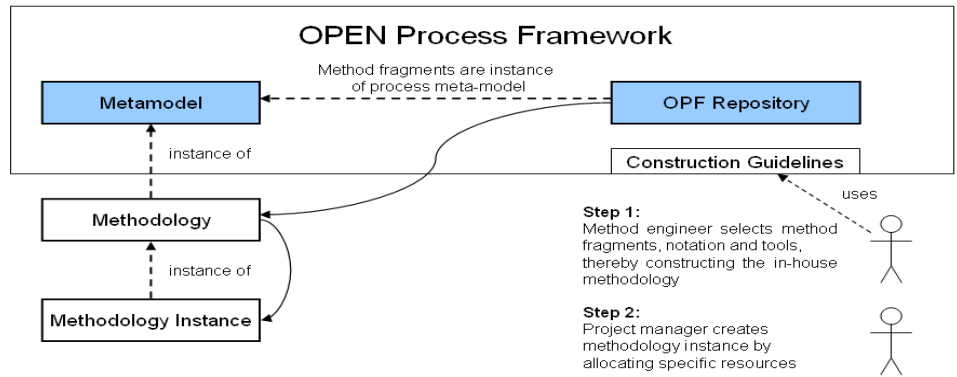

Fig.1. Construction of a project-specific SDM from OPEN's metamodel (adopted from [27]).

## *Metamodel*

The metamodel of OPEN provides a clear way for formally representing method fragments such as process models, phases, processes, tasks, techniques, work products and roles. It is imperative that each method fragment should conform to the OPEN metamodel standard. This implies that new method fragments extending the repository must be conformant with the metamodel too. It should be noted that the underpinning OPEN's metamodel has been updated and aligned with the ISO/IEC 24744 metamodel. This standard metamodel incorporates experience from earlier SME and is used to represent SDMs [25]. In this paper, we have used the recently updated OPEN metamodel terminology. Having the recent update of OPEN metamodel with ISO/IEC 24744 in mind, the five core classes of method fragments are as follows (Fig. 2) [14,15]:

1. **Work Unit**: Operations that should be performed by persons or tools to develop required Work Products. Work Units are categorized in three levels of abstraction:
    - **Process**: Process (called *Activity* in the older version of OPEN) is a coarse-grain type of typical Work Unit consisting of a cohesive collection of Tasks that produces a related set of Work Products. In other words, a Process includes a group of relevant Tasks. Sometimes, Process has been referred to as software engineering discipline.
    - **Task:** Task is a fine-grain type of Work Unit consisting of a cohesive collection of steps that produce Work Product(s).
    - **Technique:** Technique is an explicit set of procedures that explain how a Task should be performed
2. **Work Product**: Work Product is any significant produced artifact such as a diagram, a graphical or textual description, or a program produced during software development.
3. **Producer**: Persons or tools that develop expected Work Products are Producers.
4. **Language**: Language is used to represent produced artifacts using a modeling language, such as Unified Modeling Language (UML) [28], Object Modeling Language (OML) [29] or an implementation language.
5. **Stage**: Stage is intended for use in defining the overall macro-scale and time-box of a set of cohesive Work Units during the enactment of an instantiated OPEN. The whole instantiated process is structured temporally by the use of Stage concept element.

## *OPF Repository*

Besides the metamodel, OPEN contains a large number of method fragments having different levels of granularity (Processes, Tasks, and Techniques) stored in a repository. OPF recommends the use of Deontic Matrix approach [13] for selecting method fragments from repository. A two-dimensional Deontic matrix represents possible relationships between each pair of method fragments in the OPF repository. According to the five classes of OPF's method fragments, possible meaningful combinations are as follows [30]: Process/Task, Task/Technique, Producer/Task, Task/Work Product, Producer/Work Product, and Work Product/Language. For each cell of the matrix, a five-scale value can be assigned: M: Mandatory, R: Recommended, O: Optional, D: Discouraged and F: Forbidden. Processes can be considered as traditional activities in software development process such as the *Design software architecture* process, which includes a number of cohesive tasks such as *Evaluate software architecture*, *Select software architectural patterns*, *Develop initial software architecture*, *Document relevant software architecture views*, and *Realize quality attributes* [31]. To fill in the cells of the matrix with expected values, method engineer should consider many situational factors such as project size, skills of the development team, organizational culture, and usage context of the target SDM. For instance, Table 1 shows a part of decision making process of assigning enumerated values in Deontic matrix in a small B2C (business-to-customer) system [35]. Method engineer assigns possible values in order to make a mapping between Requirements Engineering process and its fine-grain requirements engineering tasks. Having situational factors of the project in mind, method engineer decides that *Identify user requirements* is mandatory for the method (denoted by M). In contrast, the Identify context task is considered as optional (denoted by O) while Conduct market research is recommended (denoted by R). It is, of course, for the other processes and tasks method fragments similar values can be assigned. OPEN metamodel and OPF repository of method fragments provide the means for SME. The OPF repository provides reusable method fragments as well as well-known and traditional processes and tasks for the construction of project-specific SDMs.

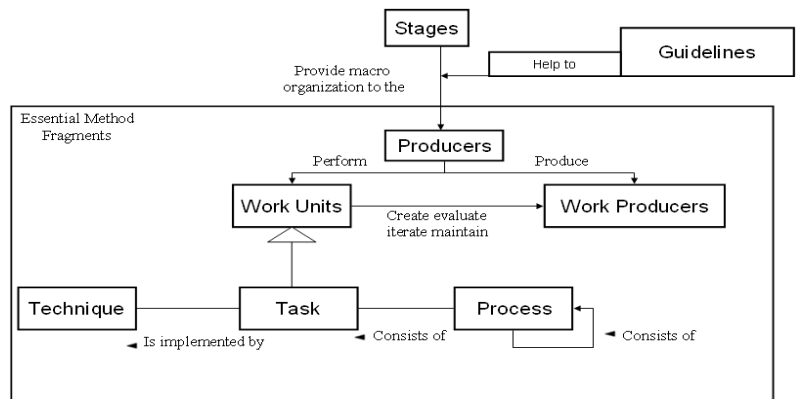

Fig.2. Constituents of OPEN's Metamodel based on ISO/IEC 24744 terminology (Adopted from [9]).

Table 1. Deontic Matrix showing the possible relations between Requirements Engineering Process and its relevant Task method fragments (adopted from [89]).

| Task | Requirements Engineering |
|---|---|
| Develop BOM | O |
| Identify context | R |
| Conduct market research | O |
| Create white site | O |
| Identify user requirements | M |
| Define problem and establish mission and objectives | O |

| Establish user DB requirements | O |

## III. RELATED WORK

OPEN has aimed to support the construction of SDMs in the manifold spectrum of software development. Over the years, several researchers have provided extensions to OPF in support of different software development approaches. Henderson-Sellers *et al.* have done significant work in enhancing OPF. They have added supportive method fragments to facilitate situational software process construction for different approaches of software development as listed below:

- *Extension for Component-Based Development (CBD) Support*: Henderson-Sellers [33] has enriched OPF by specific method fragments to support situational software process construction for component-based software development.
- *Extension for Web-Based Software Development Support***:** Concerned with characteristics of Web-Development, Haire *et al.* [34,35] have added a number of reusable method fragments to the repository for Web-based software development.
- *Extension for Aspect-Oriented Programming* (AOP) *Support*: Given that AOP aims to modularize crosscutting concerns of software development into a cohesive structure, Henderson-Sellers *et al.* [36] have added new method fragments in support of AOP to the traditional development method fragments of OPF.
- *Agent-OPEN* [37]: In this work, a number of new method fragments have been proposed to support agent-oriented software development. The OPF repository has been integrated with agent concepts. The assortment of specific agent-oriented method fragments can be found in [37].
- *Extension for Security Support:* Henderson-Sellers *et al.* [38] have presented a set of security focused method fragments have been extracted from the agent-oriented secure TROPS [40,41] methodology and added to the OPEN repository.
- Other supports for organizational transition [42,43] and usage-centered design [44] have been added to OPF too.

Although OPF has matured and contains method fragments in support of various approaches, such as OO, CBD, AOP, and Agent-OPEN development, we have identified deficiencies in the current OPF support for SOSD [45]. A major problem in SOSD arises when method engineers decide to construct a project-specific service-oriented development process. While tendency for development of service-oriented software systems and consequently appropriate service-oriented SDMs have received much attention [46,47], we investigated the current OPF method fragments and found no support for defining specific method fragments for SOSD [45]. For instance, identifying services from business processes, utilizing existing functionalities of legacy systems, discovering required services published on the Web are only a number of concerns that force to boost OPF in favor of SOSD.

Aiming at resolving the above shortcoming, we have enhanced the OPF repository with new method fragments in support of service-oriented development processes. To do this, we studied service-oriented development challenges [46] and current prominent service-oriented SDMs that prescribe successive systematic processes and tasks in order to handle service-oriented issues [48]. We have then explored service-oriented SDMs and extracted a set of processes and tasks [16] as method fragments for SOSD in conformance with the standard format of the metamodel of OPEN so that they can be easily imported into OPF tools.

# IV. SERVICE-ORIENTED SDMs: APPROPRIATE SOURCES FOR DERIVATION OF NEW METHOD FRAGMENTS

Service-orientation is currently appraised as a favorable approach in which services are utilized as fundamental elements to develop distributed software systems. Services are realized via Web–Service [50] technologies. Web-Services are independent, self-contained, reusable, and loosely coupled computational elements that form the underpinning of service-oriented systems. They collaborate via standard message protocols in a loosely coupled distributed heterogeneous environment. Therefore, a number of available published services can be composed together to develop a large software system.

In these regards, SOSD approach has emerged recently by academia, industrial practitioners and grey literature such as white papers or technical reports to address huge issues of service-oriented software systems such as *Service identification*, *Service specification*, *Service realization*, *Service discovery*, *Service composition* and *Dynamic reconfiguration*, and *Service governance* [46,47,51]. Service-oriented SDMs provide systematic processes, guidelines, and techniques required for handling of these issues. All of these end-to-end SDMs use existing traditional software engineering processes with some enhancements that are exclusive to SOSD. We briefly describe notable existing service-oriented SDMs here. The major criteria for our selection of these SDMs include their successful applications in real projects, their high maturity levels, their high rates of citations, better accessibility to their resources, and their better documentations. A comparative study of existing service-oriented SDMs can be found in [47,48,49].

- **IBM SOMA [**52**]**: In its original form, SDM included three main phases for identification, specification and realization of services. It was later improved by Arsanjani *et al.* who expanded it to seven phases including business modeling and transformation, solution management, identification, realization, specification, deployment/monitoring/management, implementation/build/assembly. SOMA is the most well-known SDM for SOSD due to its good features of software development process such as having an iterative-incremental process model, having an architecture-centric development and fractal modeling. SDM has been applied to several industrial projects successfully so that it has been designed originally from experiences of developing hundreds of real service-oriented software systems.
- **SUN SOA Repeatable Quality (RQ)** [53]: This SDM has been proposed by SUN Microsystems corporation based on Rational Unified Process (RUP) [54] and eXtreme Programming (XP) [55,56] principles that have proven mature development processes. RQ contains five phases, namely inception, elaboration, construction, transition and conception. These phases can be performed in an iterative-incremental, architecture and use-case centric development model. Applicability of RQ suffers from the lack of supportive documents describing details of internal process of SDM.
- **CBDI-SAE Process [**57**]**: This SDM is part of the CBDI-SAE SOA *Reference Framework* (*RF*) introduced by CBDI forum. It has four key phases, namely manage, consume, provide, and enable, that fully cover service-oriented development process.
- **MSOAM [**58**]**: MSOAM focuses only on service-oriented analysis and design phases. Its fully documented process prescribes systematic tasks and guidelines to develop appropriate services at different levels of granularity. However, it stops at the beginning of implementation phase.
- **IBM RUP for SOA** [59]: This iterative SDM has added service-oriented contents and specific process to RUP. In this variant of RUP, three phases of identification, specification and realization have been added.
- **SDM proposed by Papazoglou** [51]: Papazoglou *et al.* have presented a detailed service-oriented SDM that comprises eight distinct phases, namely planning, analysis and design, construction, testing, provisioning, deployment, execution and monitoring. Each phase is based on a number of principles and guidelines required for SOSD.

- **IBM SOAD (Service-Oriented Analysis Development) [60]:** SOAD's process has resulted from combining *Business Process Modeling* (*BPM*), *Object-Oriented Analysis and Design* (*OOAD*) and *Enterprise Architecture* (*EA*) practices, techniques and a number of suggested guidelines for identifying and modeling the right services. SOAD's process is rather cursory and does not fully cover service-oriented life cycle so that it would be better called as a service-oriented analysis and design technique rather than a holistic SDM. Its applicability is limited so that one can only use its specific guidelines during SOSD.
- **Service-Oriented Unified Process (SOUP) [61]:** SOUP is a hybrid SDM engineered from RUP along with XP for the development of service-oriented systems. It has six main phases, namely incept, define, design, construct, deploy, and support, in which early stages of software development look similar to those of RUP. Consequently, it has a heavyweight process and full documentation. When system becomes operational at the user environment, XP principles and practices are applied. These latter phases form a lighter process during maintenance.
- **SDM proposed by Chang and Kim [62]:** The SDM contains five phases namely identifying business processes, defining unit services, discovering services, developing services and composing services.
- **Steve Jones' Service Architectures [63]:** This SDM is based on the idea of decomposing business processes of organizations into business services resulting in business service architectures of organizations. It has a top-down view on organizations in order to get a set of business level services without their complete definition and implementation.
- **Service-Oriented Architecture Framework (SOAF) [64]:** This SDM comprises of a set of tasks, techniques, and guidelines that are grouped in five phases to address service identification and to help in deciding on service granularity while integrating existing legacy systems. Its phases are information elicitation, service identification, service definition, service realization, roadmap, and planning.

In the next section, we explain how our new method fragments have been extracted from the above enumerated service-oriented SDMs.

## V. METHODS OF IDENTIFYING RESUABLE METHOD FRAGMENTS

There are two main alternative approaches for constructing method fragments [65]: *existing method re-engineering* and *ad-hoc construction*. The first approach focuses on identifying and using method fragments from existing SDMs in a plug-and-play format. However, the ad-hoc construction approach uses real industrial projects to construct method fragments when there is no explicit defined SDM. In the latter approach, constructed method fragments are evaluated in practice; they are considered as reusable method fragments when quality standards are satisfied. Constructed method fragments in both approaches can be added to the repository of method fragments. Existing method re-engineering approach is a suitable approach to obtain method fragments when method fragments can be extracted directly from existing proven and matured SDMs. If an SDM has a successful profile in industrial realm it is reasonable to select it as a candidate and use it for constructing method fragments. Therefore, given the existence of many SDMs in the domain of SOSD, we thought it is reasonable to use them as our main source to construct our new method fragments.

By having these enumerated SDMs in mind, constructing new reusable method fragments are re-engineered from them. To do this we needed an explicit technique to identify method fragments from SDMs. It should be noted that each service-oriented SDM supports different process. Interestingly, most SDMs prescribe different tasks with different names and ambiguous and non-standard terminologies that are in fact similar. They have the same tasks in mind but from different viewpoints. If we consider them collectively in an abstract view, we can find out recurring tasks in their development processes. It is thus important to note that the multiplicity and similarity of tasks in service-oriented SDMs should be managed in some way during construction of method fragments to derive non-redundant and pure service-oriented related method fragments. The concept of process pattern can be utilized for this purpose. Process patterns are classes of common successful practices and recurring tasks (or process chunks) in SDMs [66]. In service-oriented SDMs, constituent development processes prescribe the same tasks but with different names too. Given the lack of any full-fledged technique for extraction of process patterns from service-oriented SDMs, we have previously developed a systematic technique for analyzing and mining existing service-oriented development SDMs in terms of meaningful process patterns (for more detail see [67]). We have proposed several strategies to help method engineers identifying process patterns. By focusing on contemporary service-oriented SDMs, we extracted a comprehensive set of process patterns that were refined and gradually completed. Process patterns were completed and fixed when analysis of SDMs identified no new process pattern as a candidate method fragment. Finally, we represented extracted process patterns in an existing standard repository of method fragments, namely the OPEN metamodel. Fig.3. shows our steps of extracting method fragments from service-oriented SDMs.

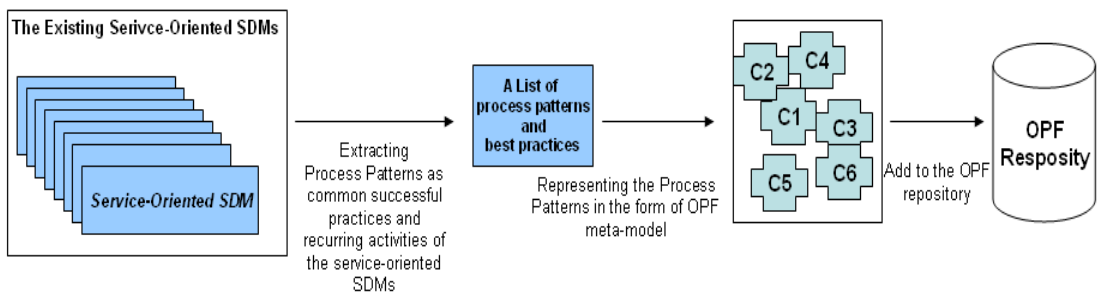

Fig.3. Existing method re-engineering approach used to obtain new method fragments

SOSD only expands existing traditional software development tasks and it is mainly considered as an evolution rather than a revolution. It can however be viewed differently as an approach to develop software out of method fragments that have semantic affinity with existing method fragments in OPF. We should avoid introducing redundant new method fragments to OPF. In other words, new formulated method fragments should be checked with existing OPF's method fragments to see if they have counterparts in OPF or not. Therefore, before considering a service-oriented method fragment as a new fragment, we checked if any method fragments existed in the OPF repository covering the new method fragment or not. If not, the method fragment was added to the repository in conformance with the OPEN metamodel standard. Although service-oriented SDMs cover traditional tasks of software development, we have discarded their related method fragments in our presentation in this paper for brevity. For instance, service-oriented SDMs emphasize on business process modeling and business process optimization and OPF supports this emphasis in the business optimization phase.

We have identified two types of method fragments that are specified in detail in the next section. *Enhanced* process method fragments enhance existing process method fragments in OPF with new specific service-oriented task(s) fragments. *New* process method fragments have not been supported by OPF and are new to this framework. Each task is described in terms of five items as follows [9]: task name, explanation, producer, work products, supportive techniques and relations. Relations specify relevant predecessor and subsequent of a task as well as Deontic matrix that was described in Section II to denote the relation of a task with other tasks, producers, techniques, and work products. Each relation can be mandatory, recommended, or optional.

# VI. PROPOSED SERVICE-ORIENTED METHOD FRAGMENTS

In this section we elaborate additional method fragments in term of processes and tasks that we propose need to be added to OPF repository to facilitate service-oriented SDM construction. Each OPF process method fragment that is enhanced with new tasks is denoted as an enhanced process while unchanged processes are not described in this paper for brevity. Each task method fragment is presented in terms of task name, a summary of the intent of the task, involved producer of the task, relevant work products, and supportive techniques.

## *1. Enhanced Process: Requirements Engineering*

In this process, the requirements of the target software system are elicited, specified and validated by all system stakeholders. This process is very similar to traditional requirements engineering. In fact, it is covered by the existing *Requirement Engineering* method fragment process in OPF. This is why we consider it as an enhanced process and hence do not explain it again. Only its difference with the *Specify Service Level Agreement (SLA)* task is described. The task is added to the Requirement Engineering process.

**Task Name:** *Specify Service Level Agreement (SLA)*

**Description:** *Quality of Services* (*QoS*) as a subset of non-functional requirements plays an important role in the service-oriented context. It forces service providers to improve their ability to meet service consumer requirements in a competitive manner with other service providers. Based on the nature of service-orientation, various service providers may provide the same service to fulfill consumer's requirements. They can however be different in QoS they provide. SLA grants the service consumer a degree of guarantee that the service provider complies with and provides acceptable QoS such as security, availability, performance, reliability in the execution environment. In this task, a contract between service providers and service consumers is established. For instance, it may be contracted that service should respond to input requests only in 20 milliseconds or less.

**Producer:** Service provider, Service consumer, Requirement engineer

**Work Products:** Document of Service Level Agreement contract

**Supportive Techniques:**

*Create SLA contract*: A consensus contracted between service consumer and service provider as *Service Level Specification* (*SLS*) document that specifies a set of typical technical parameters such as the ones listed below:

- *Purpose*: The intention of creation of the SLA contract.
- *Parties*: The service consumer and provider involved in the SLA and their responsibilities.
- *Validity Time*: The period of time that SLA should be met.
- *Scope*: The boundaries of and the expectations from SLA.
- *Service-Level Objectives*: Level of service quality that service provider and consumer agree on including service availability, security constraints, reliability, latency, and recovery time that are mostly noted in measurable and quantifiable terms.
- *Penalties*: Determining what penalties for failure must be paid when SLA contract parties violate the agreements. For instance, non-performance may be costly.

After SLA is contracted, both service provider and service consumer undertake to perform it at runtime in Web Service invocations. Table 2 shows Specify Service Level Agreement task method fragment. Third column represents the most recommended values for method fragment.

## 2. Enhanced Process: Environments Engineering

This process has many relevant tasks for assessment of the environment but it is more critical in the context of SOSD. Therefore, in this process, the status of existing infrastructure of enterprise, B2B or Systems-of-Systems (we refer to as environments) is assessed to find out candidate services from existing assets and to evaluate the readiness and existing capabilities of environment to migrate to service-oriented solution. Moreover, the reasons for migration to service-orientation are justified. This process is enhanced with the evaluate environment readiness task.

Table 2. Possible relation values of Specify Service Level Agreement task

| Elements of method fragment | Type of element |
|---|---|
| Specify Service Level Agreement (SLA) | Task |
| Service Consumer | Producer |
| Service Provider | Producer |
| Requirement Engineer | Producer |
| Create SLA Contract | Technique |
| Document of Service Level Agreement Contract | Work product |

**Task Name:** *Evaluate Environment Readiness*

**Description:** This task evaluates the readiness of environment to migrate to service-oriented solution. This task contains the following sub-tasks: evaluating the quality of existing codes and software components, evaluating reusability of valuable existing business logics of existing legacy software to expose as Web Service, evaluating quality of correctness and integrity of stored data in databases, reconstructing the architecture of existing legacy systems, and providing technology infrastructure and hardware/software resources to support secure, interoperable, reliable messaging protocols between services. Even people's attitude towards changes to their environment should be checked to find out if it is feasible to build a service-oriented solution or not.

**Producer:** Requirement engineer, Database administrator, Network administrator

**Work Products:** Report of readiness assessment

**Supportive Techniques:**

*Create a readiness report*: Requirement engineer can perfrom this task by using well-known criteria of SOA maturity models such as those proposed by IBM SOAMM and SIMM [68,69]. Table 3 shows the possible relation values of Evaluate Environment Readiness task.

Table 3. Possible relation values of Evaluate Environment Readiness task

| Elements of method fragment | Type of element |
|---|---|
| Evaluate Environment Readiness | Task |
| Requirement Engineer | Producer |
| Database Administrator | Producer |
| Network Administrator | Producer |
| Create a Readiness Report | Technique |
| Report of Readiness Assessment | Work product |

## 3. Enhaced New Process: Plan Project

The aim of this process is to perform preliminary project planning such as scheduling, risk management, and resource planning. This process does not differ from traditional project planning except in plan transition. The process includes one additional task, namely the plan transition task.

**Task Name:** *Plan Transition*
**Description:** This task is performed to adopt various strategies based on situations of the envronment and the state of existing legacy systems (software components) for transition to service-orientation [70]:
- *Replacement Strategy*: In this strategy, existing legacy systems are retired entirely by rewriting them from scratch and constructing a new service-oriented system. Although this strategy can be expensive and time consuming, it can lead to a solution that fits better to the requirements of service consumer.
- *Wrapping Strategy*: Some parts of existing valuable business logics of legacy systems are wrapped by Web Service technology (e.g., .Net or J2EE), and then exposed as a service to consumers.
- *Redevelopment Strategy*: This strategy uses the reengineering approach to add new services to existing legacy systems.
- *Migration Strategy*: This strategy incorporates both redevelopment and wrapping, and aims to develop a new system with an improved service-oriented solution.

Having selected the strategies, a transition plan that preserves functionalities of the original system for migration to service-orientation is developed. Several strategies may be pursued at the same time based on the situation of exsitnig systems. The task finishes with a primary estimation effort, cost and definition of a roadmap for migration to service-orientation. It should be noted that the plan can be updated any time.
**Producer:** Service consumer, Service provider, Project manager
**Work Products:** Transition plan, List of transition issues, Cost and effort of selected strategies
**Supportive Techniques:**
*Make Transition Plan*: The purpose of the proposed technique is to make a document in which possible alternatives for migration to service-orientation are clarified, discussed, justified, and critical milestones scheduled and documented. Table 4 shows the possible relation values of Plan Transition task.

Table 4. Possible relation values of Plan Transition task

| Elements of method fragment | Type of element |
|---|---|
| Plan Transition | Task |
| Service Consumer | Producer |
| Service Provider | Producer |
| Project Manager | Producer |
| Make Transition Plan | Technique |
| Transition Plan | Work product |
| List of Transition Issues | Work product |
| Cost and Effort of Selected Strategies | Work product |

## *4. New Process: Develop SOA Governance Model*

In this new process, a governance model is established and then cuts through all development process. Because service-orientation involves various service providers and consumers that may work in a geographically distributed environmet, a governace model should be estabilsihed to ensure that the adoption of service-orientation are constantly aligned with IT initiatives and business needs. Indeed the process acts as an umbrella process over software development that is performed continuously. This new process includes only one task.

**Task Name:** *Develop Governance Model for Current Iteration*

**Description:** Service consumers and providers collaborate to establish chains of responsibility, authority, communication, and overall scope as well as solution size and funding for performing the governance model in current iteration of the solution. Details of governance model mainly include a set of supportive high-level policies and rules to achieve right services that essentially relate to QoS. Executive mechanisms are defined to realize the defined policies. Finally, the task defines as much as possible quantifiable metrics and indicators to measure and monitor QoS during service usage.

**Producer:** Service consumer, Service provider, Requirements engineer

**Work Products:** Documented (textural description) governance model, policies, executive mechanisms, quality indicators and measurement metrics.

**Supportive Techniques:**

*Create Governance Model:* There are many techniques that service consumer and provider can accomodate as a governance model to develope service-oriented software succefully such as the one proposed by IBM [71]. Table 5 shows the possible relation values of Develop Governace Model for Current Iteration task.

Table 5. Possible relation values of Develop Governance Model for the Current Iteration task

| Elements of method fragment | Type of element |
|---|---|
| Develop SOA Governance Model | Process |
| Develop Governance Model for Current Iteration | Task |
| Service Consumer | Producer |
| Service Provider | Producer |
| Requirements Engineer | Producer |
| Create Governance Model | Technique |
| Documented (Textural Description) Governance Model | Work product |
| Policies | Work product |
| Executive Mechanisms | Work product |
| Quality Indicators and Measurement Metrics | Work product |

## 5. New Process: Design Services

The Design Services process is the core of SOSD. When the main business processes are identified and re-engineered, useful services that encapsulate business logic capabilities are defined. This process takes a set of business process models as input and yields a set of candidate services as output. The process has four tasks.

**Task Name:** *Identify Services*

**Description:** In this task, existing business processes and sub-processes are translated (manually, semi-automatically or full-automatically) into one or more services to be exposed to business partners. In other words, valuable services aligned with IT initiatives are identified. This results in a blueprint (big-picture) of service-oriented environment [63]. Definitions of identified services are more high-level and abstract than the specific details of the service operations that are rigorously specified later in the *Specify Details of Services* task.

**Producer:** Service designer (service modeler) as a member of service provider

**Work Products:** Service models, Services interfaces signatures

**Supportive Techniques:** There are three well-known techniques (typically refered to as strategy) for service identifiaction, namely [52,58]: *top-down*, *bottom-up* and *meet-in-the middle* (agile). In the top-down technique, a preliminary set of service interfaces become candidate and grouped into a logical context and further elaborated in the *Specify details of the services* task. Specifically, the technique focuses on identifying candidate services such as business services from the environment of business process models. The steps of business processes are transformed to a set of candidate services. The bottom-up technique concentrates on wrapping the underlying existing legacy logics into services that are built on top of legacy systems to make them easily accessible to other systems. This technique redirects the enviroment to new ways of supporting business needs. The agile technique proposes a combination of top-down and bottom-up techniques. Services can be modeled and presented by UML 2.0 profile for SOSD[72]. Table 6 shows the possible relation values of Identify Services task.

Table 6. Possible relation values of Identify Services task

| Elements of method fragment | Type of element |
|---|---|
| Design Services | Process |
| Identify Services | Task |
| Service Designer | Producer |
| Top-Down | Technique |
| Bottom-Up | Technique |
| Meet-In-The Middle | Technique |
| Service Models | Work product |
| Services Interfaces Signatures | Work product |

**Task Name :** *Specify Details of Services*

**Description:** The definition of defined services are consolidated with more specific details such as interface specification, service dependencies, operation signatures, operation parameters and parameter types, and input/output messages.

**Producer:** Service designer (service modeler)

**Work Products:** Service interfaces specifications, Realizer components, Service dependencies

**Supportive Techniques:**

*Add Specific Details to Services*: Service designer refines candidate services. They design interfaces to provide interoperability between service providers and consumers, input and output parameters, and error messages for services operations. Operations of services are detailed via analyzing collaborations between services. Instantiation of UML 2.0 class, interface and collaboration stereotypes [72] are used to represent services specifications. Service designer looks for potential software components that can realize service functionalities. Table 7 shows the possible relation values of Specify Details of Services task.

Table 7. Possible relation values of Specify Details of Services task

| Elements of method fragment | Type of element |
|---|---|
| Design Services | Process |
| Specify Details of Services | Task |
| Service Designer | Producer |
| Add Specific Details to Service | Technique |
| Service Interfaces Signatures | Work product |
| Software Components Specification | Work product |
| Service Dependency | Work product |

**Task Name:** *Classify Services*

**Description:** In this task, various types of identified services are classified based on the usage context. The most well-known manageable classification for services is typically hierarchical in which services are classified based on the degree of granularity from coarse-grain to fine-grain services, e.g., mission-aligned business services, enterprise services, application services and utility (also named infrastructure) services. The intent of performing the task is to facilitate clear, precise, and non-overlapping definitions for the wide range of services in the environment and may be used during a service-orientation initiative. The classification assists service providers (developers) to have more effective communication with service consumers to understand their state of existing assets and derive a blueprint for the service-oriented environment.

**Producer:** Service designer

**Work Products:** Classified service models presented by UML 2.0 stereotypes for service classification

**Supportive Techniques:**

*Classify Service*: This technique is performed to classify services based on their objectives and characteristics, e.g. business services, application services, utility services. The classification helps service providers and service consumers to identify which services will be used in the SOA layers [63]. Table 8 shows the possible relation values of Classify Services task.

Table 8. Possible relation values of Classify Services task

| Elements of method fragment | Type of element |
|---|---|
| Design Services | Process |
| Classify Services | Task |
| Service Designer | Producer |
| Classify Service | Technique |
| Classified Service Model | Work product |

**Task Name:** *Evaluate Quality of Designed Services*

**Description:** The aim of this task is to increase maintenance, simplicity, changeability, future enhancements and reuse of services. More precisely, the design quality of services is evaluated in terms of *Granularity*, *Coupling*, *Cohesion* and support of *Reusability*. The number or scope of functionalities of a service is named service granularity and is identified as a coarse-grain or a fine-grain service [51]. The appropriate level of service granularity has direct effect on service coupling and cohesion. Evaluating the coupling of services is performed to minimize dependency (e.g., data dependency and resource dependency) between services. While business processes are realized via orchestration of services, the dependency between services should be low as much as possible to provide more agility of business processes while underlying business processes and rules change more frequently upon business needs. Low coupling increases service reusability for future projects. Evaluation of cohesion is performed to check whether a service exposes a set of relatedness of necessary functionalities or not. It should be noted that a tradeoff needs to be made while taking into account granularity, coupling, cohesion, and service reusability.

**Producer:** Service desginer

**Work Products:** Refined service model

**Supportive Techniques:** There are three specific service-oriented techniques for performing this task.

*Evaluate Service Granularity:* Service granularity can be evaluated in different ways such as by the number of software component interfaces invoked for a given service operation [64]. When service operations increase, the sizes of messages and data transfers increase and create higher dependency on the context. In contrast, fine-grain services increase the number of message passing between them.

*Evaluate Service Coupling:* Service designer utilizes the techniques such as the one proposed by Perepletchikov *et al.* [73] in which a suite of seventeen quantified service coupling metrics are proposed to measure service coupling. Based on the evaluation results, service modeler may revise the service model. Furthermore, prescriptive guidelines can be incorporated during service design [46].

*Evaluate Service Cohesion*: Service designer can use this technique to determine if the functionalities of a designed service are cohesive for example if coincidentally and sequentially of operations of Web Services are satisfied or not [74]. Table 9 shows the possible relation values of Evaluate Quality of Designed Services task.

Table 9. Possible relation values of Evaluate Quality of Designed Services task

| Elements of method fragment | Type of element |
|---|---|
| Design Services | Process |
| Evaluate Quality of Designed Services | Task |
| Service Designer | Producer |
| Evaluate Service Granularity | Technique |
| Evaluate Service Coupling | Technique |
| Evaluate Service Cohesion | Technique |
| Refined Service Model | Work product |

## *6. Enhanced Process: Service-Oriented Architecture Engineering*

This process is supported by existing *Architecture Engineering* process in the OPF repository that we renamed it to *Service-Oriented Architecture Engineering* to promote it to SOA. The main enhancement relates to instantiation of stack-based service-oriented reference architecture (SOA reference model) [52] in which services are organized into different layers. The layered architecture enables complexity management and facilitates the decision to where to place services and how to provide support for SOA-specific QoS issues. QoS is realized by utilizing well-known architecture strategies and tactics such as the ones proposed in [75].

## *7. Enhaced Process: Develop Services*

This process enhances the *Implementation* process of the OPF repositoy. The real required services such as business services, enterprise services, application services and utility are developed in various manners. The process includes three tasks as follows.

**Task Name:** *Implement and Test Necessary Services*
**Description:** If no suitable required Web Service is found in OPF or no exact match with the requirements is found, an alternative implementation must be developed from scratch. Services are implemented and tested by service provider (development team). Meanwhile, specification of the implemented service as a Web Service is expressed in Web Service Description Language (WSDL) wherein public available operations are exposed in a way that service consumers can invoke them. Because Web Services can be developed separately by geographically distributed development team, all Web Services as part of a distributed system should be tested independently and integrated with other Web Services or systems involved. Test performed by service-provider and service consumer. Service provider can provide a number of test cases for service consumer to reuse.
**Producer:** Service developer , Service tester
**Work Products:** Executable Web Services, Services WSDLs and WS-Policy
**Supportive Technique:**

*Implement Services*: Service developer uses this technique. Existing OO analysis and design techniques such as analyzing and designing classes, CRC card modeling and classifying relevant classes into cohesive software components are used to implement services. From an implementation viewpoint, a Web Service realizes a service comprising of a number of software components. Specifications of software components provide the basis for the design and implementation of Web Services, i.e. service interfaces. OPEN has a set of method fragments that allows for incorporating CBD approach in the software developement process. The Implement Services task forces service providers to accomodate existing tasks of OPF that are specified in the Implementation process.

*Perform WSDL Testing*: In addition to traditional testing techniques, Service tester can use the WSDL testing technique. Web Services have WSDL as the only available interface at testing time. WSDL metadata files are XML documents containing information about Web Service's operations and required QoSs. Test-case generator tools use WSDL files to generate test cases automatically. Test cases act as SOAP messages sent to Web Services as well as to service consumers. All Web Service operations include various inputs/outputs with different data types. Confidentiality and integrity of SOAP messages during test should be taken into account too. Table 10 shows the possible relation values of Implement and Test Necessary Services task.

Table 10. Possible relation values of Implement and Test Necessary Services task

| Elements of method fragment | Type of element |
|---|---|
| Implement and Test Necessary Services | Task |
| Service Developer | Producer |
| Service Tester | Producer |
| Implement Services | Technique |
| Perform WSDL Testing | Technique |
| Executable Web Services | Work product |
| Services WSDLs and WS-Policy | Work product |

**Task Name:** *Implement Necessary Wrappers*

**Description:** This task concentrates on the software components comprising the interfaces of existing legacy systems. Based on the work products of the Evaluate environment readiness task, valuable business logics of one or more existing legacy systems that provide desired functionalities are extracted and exposed through universal standard Web Service technologies such as .Net or J2EE. Wrapping provides new broad accessibility Web Service interfaces to existing legacy software components. Wrapping existing software components interfaces as Web Services is justifiable when the development of service-oriented systems from scratch is expensive, risky and time consuming.

**Producer:** Service developer

**Work Products:** Executable Web Services, Services WSDLs

**Supportive Techniques:**

*Implement Wrappers*: There are several step-by-step techniques including manual [76,77], semi-automaticlly [78,79], or fully-automatically [80] techniques that service developers can use to wrapp individual functionalities in legacy source codes such as Web Services. Table 11 shows the possible relation values of Implement Necessary Wrappers task.

Table 11. Possible relation values of Implement Necessary Wrappers task

| Elements of method fragment | Type of element |
|---|---|
| Implement Necessary Wrappers | Task |
| Service Developer | Producer |
| Service Tester | Producer |

| Implement Wrapper | Technique |
| --- | --- |
| Executable Web Services | Work product |
| Services WSDLs | Work product |

**Task Name:** *Develop Necessary Composite Web Services*

**Description:** This task composes of a number of prepared fine-grain (also called atomic) Web Services and other software components related to the underlying business processes that form a more coarse-grain Web Service called a Composite Web Service that is assumed to maximize business value. In fact, service consumers synthesize composite Web Services to realize ultra large-scale software system in terms of Systems-of-Systems or supply chain management via composing a dozen of heterogeneous distributed independent Web Services. Definition of business service is adopted from Business Process Modeling Language (BPML) [81] and IBM's WSFL [82] wherein the invocation order of Web Services - orchestration or choreography- and the sequencing of message passing and bindings between services are defined to form flow of business services. It is worth to note that the composition of appropriate Web Services with guaranteed QoS should be considered prior to the construction of a service-oriented system. Therefore, analysis of Web Service composition alternatives to select the best composition must be done. Composite Web Services are executed later by Business Process Execution Language for Web Services (BPEL4WS) [83], which is a standard engine for business process execution. The task is conducted using manual, semi-automated, or automated composition techniques [84].

**Producer:** Service consumer, Business process engineer

**Work Products:** Composite services as business process (BPEL processes)

**Supportive Techniques:**

*Compose Web Service*: There are several supportive techniques for this task [85]. A service consumer takes a number of fine-grain Web Services to configure a given business process model. Then he/she evaluates how best the composed Web Services meet the desired functionalities and QoS parameters to select the best composition. Service composition task becomes more complex as the number of provided Web Services (e.g., available Web Services on the Web) increases. Therefore, automated or semi-automated tools to help service consumers in this hard task are critically required. Table 12 shows the possible relation values of Develop Necessary Composite Web Services task.

Table 12. Possible relation values of Develop Necessary Composite Web Services task

| **Elements of method fragment** | **Type of element** |
| --- | --- |
| Develop Necessary Composite Web Services | Task |
| Service Consumer | Producer |
| Business Process Engineer | Producer |
| Compose Web Services | Technique |
| Composite Services as Business Process | Work product |

## *8. Enhanced Process: Reuse Engineering*

We have only enhanced this existing OPF process with one service-oriented specific task.

**Task Name:** *Discover Necessary Web Services*

**Description:** The aim of this task is to help searching for and selecting from existing Web Services that match best with service consumers' requirements such as QoS and functionalities. The result of this task is a list of retrieved candidate Web Services. Given that Web Services can be developed by various service providers, services should be certified to ensure that selected services satisfy the required quality of concerns (SLA). It is possible that many Web Services exactly match the particular requirements. Therefore, service consumer must evaluate them all to select the best ones. For paid services, a usage-based billing model for charging of services is contracted between service provider and service consumer. Typically, the discovery task is supported by automatic Web Service discovery engines.

**Producer:** Service consumer
**Work Products:** Executable Web Services
**Supportive Techniques:**
*Search Web Services*: Service consumers can use generic search engines such as Google to find WSDL documents in the Web or running SOAP APIs that allow performing queries on UDDI directories. Tools can assist service consumers to accurately locate services that satisfy the required QoSs. Table 13 shows the possible relation values of Discover Necessary Web Services task.

Table 13. Possible relation values of Discover Necessary Web Services task

| Elements of method fragment | Type of element |
|---|---|
| Discover Necessary Web Services | Task |
| Service Consumer | Producer |
| Search Web Services | Technique |
| Executable Web Services | Work Product |

## 9. Enhanced Process: Enable Service-Oriented Solution

This process is mainly supported by *Deployment* process in OPF. The *Service-Oriented Solution* enhances this process by two service-oriented specific tasks. In this process, Web Services are deployed in an operational environment. Moreover, defects and missing requirements are discovered in this process. In some cases, it is difficult to determine a time for deployment of Web Services as building blocks of the system when a service-oriented system can be fully developed via existing Web Services that have already been provided and published by various service providers.

**Task Name :** *Publish Web Services*
**Description:** Web Services are hosted and advertised by service providers and published to an accessible common registry such as a Universal Description Discovery and Integration (UDDI) server [86]. The major information in addition to what is provided for a typical Web Service includes Web Service's operations signatures and QoS values such as the cost of usage, availability, and security issues. Service consumer can discover the required Web Services through universal protocols such as SOAP messages. In fact, service providers advertise their Web Services at a global market place on the Web.
**Producer:** Service installer
**Work Products:** Deployed and published services
**Supportive Techniques:**
*Import Web Services into a Common Web Service Repository*: The service installer takes tested Web Services, generates a Web Service description document like WSDL for each one, and publishes it to a common repository such as in a UDDI where service consumers can find Web Services. Table 14 shows the possible relation values of Publish Web Services task.

Table 14. Possible relation values of Publish Web Services task

| Elements of method fragment | Type of element |
|---|---|
| Publish Web Services | Task |
| Service Installer | Producer |
| Import Web Services into the Common Web Service Repository | Technique |
| Deployed and Published Services | Work Product |

**Task Name:** *Perform Test in Large*

**Description:** This task tests orchestrated or choreographed Web Services to see if the composition of Web Services that build a distributed system actually meet the business acceptance criteria for functional requirements and SLA for nonfunctional concerns. Based on the nature of SOSD, such a test typically involves more than one software development team (service provider) and business partner (service consumer) such as when a composite Web Service realizes a Business-to-Business (B2B) business process.

**Producer:** Orchestrator/Choreographer Tester

**Work Products:** Test cases, Result of running test cases

**Supportive Techniques:**

*Perform Orchestration/Choreography Testing*: One way to perform this task is to define certain business process scenarios as test cases. The results of performing the tests are compared with expected functionalities, SLA contracts, specially predefined policies, and quality criteria in the SOA governance criteria. Table 15 shows the possible relation values of Perform Test in Large task.

Table 15. Possible relation values of Perform Test in Large task

| Elements of method fragment | Type of element |
|---|---|
| Perform Test in Large | Task |
| Orchestrator/Choreographer Tester | Producer |
| Perform Orchestration/Choreography Testing | Technique |
| Test Cases | Work Product |
| Results of Running Test Cases | Work Product |

## *10. Enhanced Process: Maintenance*

After Web Services are fully deployed in an operational environment, this process evaluates QoS of all participating Web Services that make up the distributed system, against predefined SLA contract and SOA governance model continuously. This process includes one main task.

**Task Name:** *Monitor Operational Web Services*

**Description:** The aim of this task is to indicate service degradation, noncompliance with service-level offerings, and service availability levels before service failure actually occur. To do this, service consumers gather and log data during Web Services usage. They then measure and interpret Web Services against predefined metrics in *Develop SOA Governance* model and the SLA contract. For Web Services having usage-based billing models as well as those consensued in the SLA contract, service proivders generate billing reports to service consumers to pay them.

**Producer:** Service consumer, Service provider

**Work Products:** Statically-generated reports of QoS, Service metering, Billing report and Defect report.

**Supportive Techniques:**

*Monitor QoS of Web Services*: Service consumers log and analyze Web Service invocations, for instance the number of Web Service operation invocations or the number of authentication failures, to detect violations from promised QoS parameters such as response time, throughput, and availability. Historical information of Web Services' QoSs are analyzed. Based on the generated reports, business processes may need management decisions to accommodate changes to business processes (as composite Web Services) such as Web Service replacement with another one for continuous QoS improvement (See the *Compose Web Service Dynamically* task). Table 16 shows the possible relation values of Monitor Operational Web Services task.

Table 16. Possible relation values of Monitor Operational Web Services task

| Elements of method fragment | Type of element |
|---|---|
| Monitor Operational Web Services | Task |
| Service Consumer | Producer |
| Service Provider | Producer |
| Monitor QoS of Web Services | Technique |
| Statically Reports of QoS | Work Product |
| Service Metering | Work Product |
| Billing Report and Defect Report | Work Product |

**Task Name:** *Compose Web Services Dynamically*

**Description:** This task allows utilizing various Web Services on the Web on demand without enforcing any Web Service composition or deployment in advance. This way, service composition that is usually performed at design-time can thus be done dynamically at runtime too. Consequently, this task blurs the distinction between tasks at design time and runtime. Malfunctioning of Web Services at runtime in the system can be sensed and rectified dynamically by probing for new Web Services and replacing them with new ones on the fly.

**Producer:** Service consumer

**Work Products:** New Discovered Web Services

**Supportive Techniques:**

*Reconfigure Composite Web Services*: Service consumers reconfigure composite Web Services in which degraded Web Services have been detected and replaced with new ones. Typically, dynamic Web Service composition is performed automatically. Table 17 shows the possible relation values of Compose Web Service Dynamically task.

Table 17. Possible relation values of Compose Web Service Dynamically task

| Elements of method fragment | Type of element |
|---|---|
| Compose Web Services Dynamically | Task |
| Service Consumer | Producer |
| Reconfigure Composite Web Services | Producer |
| New Discovered Web Services | Technique |

**Relation among method fragments:** Although method fragments are stored independently in the repository, the constraints on them can be specified via Constraint super-class of OPEN [14,15]. Constraint super-class provides a linkage as well as predecessor and subsequent for method fragments. There are two subtypes of Pre-Condition and Post-Condition for this purpose that we have used. The constraints allow us to clarify imperative constraints on method fragments. Fig. 4 shows possible predecessor and subsequent constraints as pre-conditions and post-conditions of the use of task method fragments as recommended in OPEN. Directions of arrows show dependency between tasks For instance, Identify Services task should be completed for identifying a list of required services before performing the Discover Necessary Web Services for obtaining executable Web Services. It is obvious that all constraints are maintained as Pre-Condition and Post-Condition fields of the task method fragments.

## VII. POSITION OF NEW METHOD FRAGMENTS IN OPEN

Our proposed method fragments represent necessary service-oriented method fragments that should be added to the OPF repository in support of service-oriented SDM construction using OPF method fragments. To construct a project-specific service-oriented SDM, required process should be selected first. Then task method fragments should be selected to complete the internal details of process of method fragments. For each task method fragment, relevant producer(s), work product(s) and supportive technique(s) should be determined.

Table 18 shows the position of the new service-oriented method fragments in the OPEN process model as an enhancement to the OPF repository in order to incorporate service-oriented method fragments. The original process method fragments (first column) of OPEN with 10 process form the OPEN development process model. New process and task method fragments enhance OPEN in portions that service-oriented support is needed. The two new processes of method fragments are Design Services and Develop Governance. Each of these new processes has new task method fragments themselves. The original processes of method fragments are extended with new service-oriented task method fragments (second column). For instance, the Requirements Engineering process is enhanced with Specify SLA task method fragment. For brevity, task method fragments originally existing in OPF are not shown here (for details see [14,15]). The third column shows the Producer that should be employed to produce necessary Work Products (forth column). Tasks are performed to complete the processes. Supportive techniques should be used to realize tasks. For instance, the Design Services process has four associated tasks.

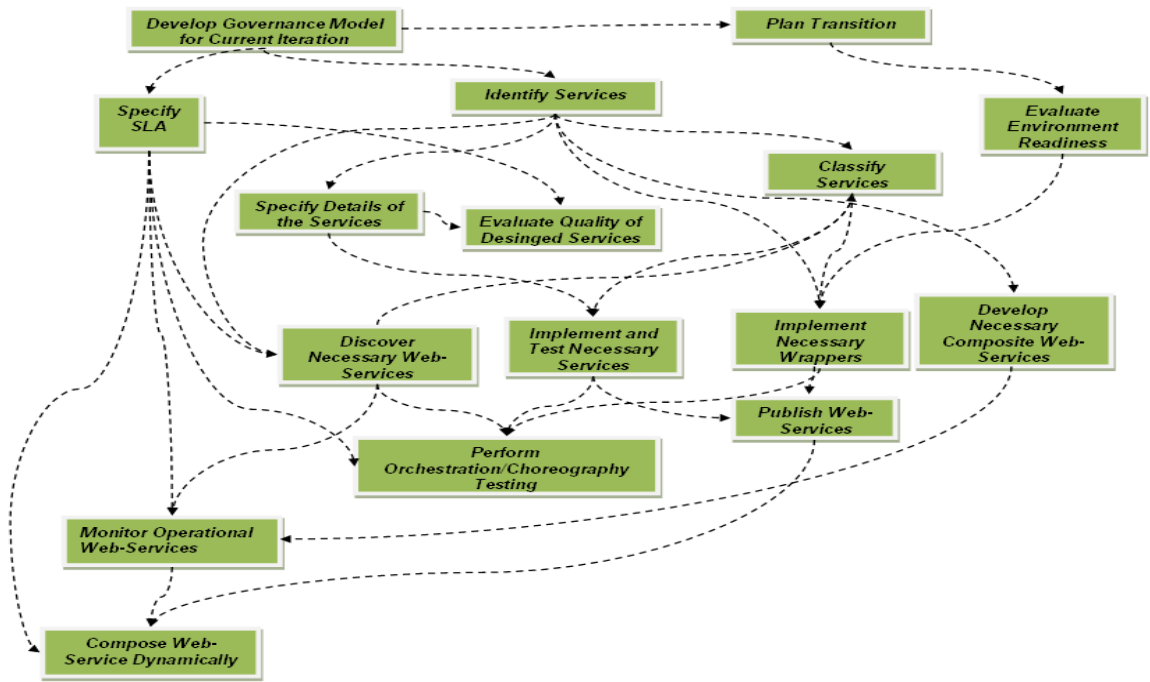

Fig.4. Relation among method fragments

Table 18. New service-oriented specific method fragments incorporated into OPEN

| Process | Task | Technique | Work Product | Producer |
| --- | --- | --- | --- | --- |
| Requirements Engineering | 1.Specify SLA | a. Create SLA Contract | a. Document of Service Level Agreement Contract | a. Service Provider, b. Service Consumer, c. Requirement Engineer |
| Environments Engineering | 1. Evaluate Environment Readiness | a. Evaluate Environment with SOA Maturity Model Criteria | a. Report of Readiness Assessment | a. Requirement Engineer, b. Database Administrator, c. Network Administrator |
| Develop Governance | 1.Develop Governance for Current Iteration | a. Create Governance Model | a. Documented Governance Model b. Policies, c. Executive Mechanisms, | a. Service Consumer, b. Service Provider, c. Requirements Engineer |

|  |  |  | d. Quality Indicators and Measurement Metrics |  |
|---|---|---|---|---|
| Reuse Engineering | 1. Discover Necessary Web Services | a. Search Web Services | a. Executable Web Services | a. Service Consumer |
| Design Services | 1. Identify Services | a. Top-Down Analysis b. Bottom-Up Analysis c. Meet-In-the Middle Analysis | a. Service Models b. Service Interface Signatures | a. Service Designer |
|  | 2. Specify Details of Services | a. Add Specific Details to the Service | a. Service Interface Signatures b. Realizer Components c. Service Dependency | a. Service Designer |
|  | 3. Classify Services | 1.Classify Service | a. Classified Service Model | a. Service Designer |
|  | 4. Evaluate Quality of Designed Services | a. Evaluate Service Granularity b. Evaluate Service Coupling c. Evaluate Service Cohesion | a. Refined Service Model | a. Service designer |
| Implementation | 1. Implement and Test Necessary Services | a. Implement Services b. Perform WSDL Testing | a. Executable Web Services | a. Service Developer b. Service Tester |
|  | 2. Implement Necessary Wrappers | a. Implement Wrappers | a. Services WSDLs and WS-Policy | a. Service Developer |
|  | 3. Develop Necessary Composite Web Services | a. Compose Web Service | a. Executable Web Services b. Services WSDLs c. Composite Services as Business Process | a. Service Consumer b. Business Process Engineer |
| Deployment | 1. Publish Web Services | a. Import Web Services into the common Web Service Repository | a. Deployed and Published Services | a. Service Installer |
|  | 2. Perform Test in Large | a. Perform Orchestration or Choreography Testing | a. Test cases b. Result of running Test Cases | a. Orchestrator /Choreographer Tester |
| Maintenance | 1. Monitor Operational Web Services | a. Monitor the QoS of Web Services | a. Static Reports of QoS b. Service Metering c. Billing Report and Defect Report | a. Service Consumer b. Service Provider  a. Service Consumer |
|  | 2.Compose Web Service Dynamically | a. Reconfigure Composite Web Services | a. New Discovered Web Services | a. Service Consumer |
| Management | 1. Plan Transition | a. Make Transition Plan | a. Transition Plan b. List of Migration Issues c. Cost and Effort of Selected | a. Service Consumer, b. Service Provider, c. Project Manager |

| | | | Strategies | |
|---|---|---|---|---|
| | | | | |

# VIII . EVALUATION

We presented a set of reusable service-oriented method fragments to facilitate the construction of situational SDM methods based on situational factors of the project at hand. Whenever an organization aims to construct a service-oriented SDM, it can construct its SDM based on our proposed set. The question is though how much valid and correct are these fragments? We thus need to verify and validate (V&V) our proposed method fragments.-They are however similar to any other kind of *software artifact* in the software engineering domain such as software components [89], analysis patterns [90] and design patterns [91] that are being argued in order to increase reusability and saving construction time, there is urge to assess quality of resulting enhanced repository. In other words, it is necessary to evaluate Verification and Validation (V&V) of the proposed method fragments.- To do this, we need to state clearly, what we mean by V&V of a set of method fragments. Unfortunately, there is no pertinent and approved definition or analysis criteria to verify and validate a set of method fragments to be added to the OPF repository. Henderson-Sellers and Gonzalez have conducted a theoretical work on the granularity and the size of the resulting method fragments [92]. They have argued that granularity affects reusability of method fragments and thus method fragments should be atomic rather than being coarse grained. However, their work is in progress and not finalized yet. Therefore, we could not find a mature metric or evaluation criteria to analyze our method fragments in detail. Furthermore, most existing evaluation criteria, which often use qualitative questionnaires, have focused on evaluating the quality of the constructed situation SDM rather than the method fragments themselves [19].

For the above reasons, we decided to use the abstract definition of V&V about software artifacts proposed by Bohem's [93] and Pressman [94]:

*Abstract Verification*: Has the artifact been constructed in the right way?

*Abstract Validation*:  Has the right artifact been constructed?

Using these definitions, we made an analogy between the terminologies of V&V in the realms of software engineering and SME, specifically method fragments. To be more specific, we have concretized V&V for SME as follows:

*Concrete Verification*: Has the proposed method fragments been constructed/identified in a right manner in line with the OPEN/OPF and SME objectives?

*Concrete Validation*: Has the right method fragments been developed to facilitate the construction of various SDMs?

These definitions is analogous to V&V of any kind of produced software artifact in soft engineering area. Considering this analogy, we followed to find out fair evaluation way in which method fragments can be verified and validated.

In terms of verification of proposed method fragments, we systematically reviewed main sources and published literature about OPEN metamodel [9,14], OPF repository and construction of method fragments [19]. We met two main concerns that should be addressed adequately whenever adding new extension support to OPF repository is aimed. Firstly, as stated in [34] each new set of method fragments for adding to OPF repository must conform OPEN metamodel notions, being abstractness and technology-independence. Secondly, method fragments must be identified through a systematic technique [ref].

To address the first concern, we made sure to represent the proposed method fragments in the same structure as that of OPEN to make them consistent with the method fragments already stored in OPF and thus easily connectable to preexisting method fragments. Meanwhile, the underpinned metamodel used for method fragments is standard whilst there are manifold metamodels whereas they adopt different notions and elements for modeling method fragments. Indeed, leverage on a standard metamodel, like OPEN used in this paper, provide the multitude organization benefit a common language and better understanding for key service-oriented method fragment.

As for the second concern, we have used the method re-engineering approach proposed by Ralyté [7] and systematically reviewed the main sources and published literature on OPEN metamodel [9,14], OPF repository, and construction of method fragments [19]. We then extracted the recurring fragments from eleven prominent service-oriented SDMs to ensure the resulting method fragments are [16] non-redundant, without overlapping, and compatible with the structure of the OPEN's metamodel [25].

To validate we have developed the right method fragments, we use two criteria presented in [95,96,97,98], namely *usability* and *completeness*. The usability criterion measures the range of situational SDMs that can be constructed from the proposed method fragments based on the projects' requirements. The completeness criterion measures how fully the proposed method fragments cover any specific domain of software development. In the following two sub-sections A and B, we separately argue that the proposed method fragments satisfy these two criteria in practice.

## A. COMPLETENESS

We use *Domain Fragment*s and *Domain Coverage* presented by Han [98] to validate the completeness of our proposed method fragments. Domain coverage measures the adequacy of a set of proposed method fragments in covering a specific domain of software development while domain fragments are a subset of a domain and propitious domain fragments are those that more fully cover that domain. We thus need to define a domain for validating the completeness of our proposed method fragments first. Considering the Papazoglou's recommendation [2] that argues in favor of service-oriented SDMs as a suitable representative domain for service-oriented paradigm or service-oriented software engineering/computing, service-oriented SDMs constitute the domain of our work. Fortunately, we had in fact selected this domain before to derive the proposed method fragments.

As stated in Section IV before, we had selected a number of prominent service-oriented SDMs based on their applications in real projects, their maturity levels, their citation rates, accessibility to their resources, and quality of their documentations [49]. The following eleven service-oriented SDMs were chosen: IBM SOMA, SUN SOA Repeatable Quality (RQ), CBDI-SAE Process, MSOAM, IBM RUP for SOA, SDM proposed by Papazoglou, IBM SOAD (Service-Oriented Analysis Development), Service-Oriented Unified Process (SOUP), SDM proposed by Chang and Kim, Steve Jones' Service Architectures, Service-Oriented Architecture Framework (SOAF). Therefore, all these eleven SDMs constitute the domain of our work and the set of proposed method fragments constitute the domain fragments. We should thus show that the proposed method fragments *(i.e., domain fragments)* cover these eleven service-oriented SDMs *(i.e., domain)* adequately. We define two general equations below (Equation I and II) to measure the adequacy of this coverage, wherein

- *Task* refers to a substantial task in a SDM. It is a bit inductive and tentative to figure out which elements in a SDM are tasks. Some examples include *Requirement elicitation*, *Design prototypes*, *Evaluate software architecture*, and *Implement code*.
- *Number of Tasks (NT)* represents the total number of tasks in a SDM.
- *Method Fragment (MF)* **represent a typical method fragment.**

- *Sum of Method Fragments (SMF)* represents the total number of service-oriented task method fragments, which is sixteen in our case here in this paper.
- *N* represents the number of SDMs, which is eleven in our case here.
- *Method Coverage (MC)* represents the degree of coverage of a service-oriented SDM (a SDM is a subset of domain) by a set of service-oriented method fragments (domain fragments) that is calculated by Equation I.
- *Domain Coverage (DC)* represents the degree of coverage of the service oriented SDMs (domain) by service-oriented method fragments (domain fragments) that is calculated by Equation II.

$$MC = \frac{\sum_{i=1}^{SMF} MF_i}{\sum_{i=1}^{NT} Task_i} \begin{cases} > 1 \\ = 1 \\ < 1 \end{cases} \quad (I)$$

$$DC = \begin{cases} 1 & \forall\ MC \in domain \Rightarrow MC = 1 \\ 0 & else \end{cases} \quad (II)$$

MC refers to the point that a whether or not service-oriented method fragments can cover a specific service-oriented SDM. Three possible values may be resulted for MC. An MC greater than one means that the proposed service-oriented method fragments not only cover a SDM but that they provide more tasks than required by the SDM. In other words, the SDM can be constructed by reusing the proposed method fragments. An MC equal to one implies a one-to-one relation between method fragments and domain. An MC less than one means that method fragments are not adequate to cover the SDM fully and that they should be enriched with more method fragments. DC is one when all calculated MC values are one. In the other words, method fragments cover all SDMs. DC zero when they fall short of covering all SDMs.

Table 19 shows the calculated MC values for each SDM by the proposed method fragments, using Equation I. It should be noted that the calculation of the number of tasks in SDMs was difficult because tasks were represented mostly in a narrative form rather than in a formal format. We thus used the process-centered textual template proposed by Ramsin [99] to categorize tasks and facilitate their enumerations. So the second column of Table 19 shows the list of decomposed tasks of each SDM using this process-centered template. For brevity, we did not consider traditional tasks of SDMs in this template. For example, in the IBM RUP for SOA there were three main service-oriented tasks. The third column demonstrates the correspondence between one or more proposed service-oriented task method fragments to each task of each SDM. In other words, the second and third columns together show a one-to-one mapping between the tasks of SDMs and the proposed set of method fragments.

As it is shown in Table 19, the MC values for all eleven SDMs were lower than one. For example, given our 16 proposed method fragments (SSMF = 16), IBM SOAD with three tasks (NA=3) had an MC equal to 3/16 (0.1870) indicating that the proposed method fragments not only cover this SDM but provide more support than it required. In other words, a method engineer can construct IBM SOAD with the proposed task method fragments. The same is true for other SDMs too.

Given that the MC values for all eleven SDMs were lower than one, the DC value is one indicating that the proposed task method fragments cover *all* of SDMs, or that all these SDMs can be constructed using the proposed method fragments. We have thus shown that the proposed method fragments *(i.e., domain fragments)* cover these eleven service-oriented SDMs *(i.e., domain)* adequately, or better said are complete.

TABLE 19. COVERAGE OF ELEVEN SERVICE-ORIENTED SDMs BY THE PROPOSED SERVICE-ORIENTED METHOD FRAGMENTS

| SDM | Task | Corresponding task method fragment(s) |
|---|---|---|
| **IBM SOAD** | 1. Service Identification | Identify Services |
| | 2. Service Classification | Classify Services |
| | 3. Service Modeling and Documentation | Specify Detail of Services |
| | NA = 3    SSMF = 16    MC = 3 / 16 (0.187) | |
| | **Task** | **Corresponding task method fragment(s)** |
| | 1. Business Modeling and Transformation | Business Requirements Engineering (from Requirements Engineering process method fragment in OPF) |
| | 2. Solution Management | All tasks in Project management Process method fragments in OPF |
| | 3. Identification | Identify Services |
| | 4. Specification | Specify Detail of Services |
| **IBM SOMA 2008** | 5. Realization | Candidate Component Evaluation and Candidate Component Solution Identification in OPF (from Component Product Acquisition process method fragment in OPF repository) |
| | 6. Implementation | Implement and Test Necessary Services, Implement Necessary Wrappers |
| | 7. Deployment, Monitoring, and Management | Publish Web Services, Monitor Operational Web Services, Compose Web Services Dynamically |
| | NA = 7    SSMF = 16    MC = 7 / 16 (0.437) | |
| | **Task** | **Corresponding task method fragment(s)** |
| | 1. Manage | Evaluate Environment Readiness, Develop Governance Model for Current Iteration |
| | 2. Consume | Business Requirements Engineering (from Requirements Engineering process method fragment in OPF) |
| **CBDI-SAE Process** | 3. Provide | Plan Transition, Service-Oriented Architecture Engineering, Implement and Test Necessary Services, Implement Necessary Wrappers |
| | 4. Enable | Publish Web Services, Monitor Operational Web Services Compose Web Services Dynamically |
| | NA = 4    SSMF = 16    MC = 4 / 16 (0.25) | |
| | **Task** | **Corresponding task method fragment(s)** |
| | 1. Incept | Evaluate Environment Readiness and Business Requirements Engineering (from Requirements Engineering process method fragment in OPF) |
| | 2. Define | Plan Transition, Identify Services, All tasks in Project management Process method fragments in OPF |
| **SOUP** | 3. Design | Specify Detail of Services |
| | 4. Construct | Implement and Test Necessary Services, Implement Necessary Wrappers |
| | 5. Deploy | Publish Web Services |
| | 6. Support | Monitor Operational Web Services |
| | SMA = 6   SSMF = 16   MC = 6 / 16 (0.375) | |
| | **Task** | **Corresponding task method fragment(s)** |
| | 1. Service-Oriented Analysis | Evaluate Environment Readiness, Business Requirements Engineering (from Requirements Engineering process method fragment in OPF) |
| | 2. Service-Oriented Design | Identify Services, Service-Oriented Architecture Engineering |
| **MSOAM** | 3. Service Development | Implement and Test Necessary Services, Implement Necessary Wrappers |
| | 4. Service Testing | Implement and Test Necessary Services, Implement Necessary Wrappers |
| | 5. Service Deployment | Publish Web Services |
| | 6. Service Administration | Monitor Operational Web Services, |

| | | Compose Web Services Dynamically |
|---|---|---|
| | **NA = 6   SMF = 16    MC = 6 / 16 (0.375)** | |
| **IBM RUP for SOA** | Task | Corresponding task method fragment(s) |
| | 1. Service Identification | Identify Services |
| | 2. Service Specification | Specify Detail of Services |
| | 3. Service Realization | Candidate Component Evaluation (OPF), Candidate Component Solution Identification in OPF (from Component Product Acquisition process method fragment in OPF repository) |
| | **NA = 3   SMF = 16    MC = 3 / 16 (0.187)** | |
| **SUN SOA RQ** | Task | Corresponding task method fragment(s) |
| | 1. Inception | Evaluate Environment Readiness, Business Requirements Engineering (from Requirements Engineering process method fragment in OPF) |
| | 2. Elaboration | Evaluate Environment Readiness Service-Oriented Architecture Engineering Business Requirements Engineering (from Requirements Engineering process method fragment in OPF) |
| | 3. Construct | Implement and Test Necessary Services, Implement Necessary Wrappers |
| | 4. Transition | Publish Web Services |
| | 5. Maintenance | Monitor Operational Web Services, Compose Web Services Dynamically |
| | **NA = 5   SMF = 16    MC = 5 / 16 (0.312)** | |
| **SOAF** | Task | Corresponding task method fragment(s) |
| | 1. Information Elicitation | Business Requirements Engineering (from Requirements Engineering process method fragment in OPF) |
| | 2. Service Identification | Identify Services |
| | 3. Service Definition | Specify Detail of Services |
| | 4. Service Realization | Candidate Component Evaluation, Candidate Component Solution Identification in OPF (from Component Product Acquisition process method fragment in OPF repository) |
| | 5. Road Map and Planning | Develop Governance Model for Current Iteration, Plan Transition |
| | **NA = 5   SMF = 16    MC = 5 / 16 (0.312)** | |
| **Steve Jones' Service Architectures** | Task | Corresponding task method fragment(s) |
| | 1. Initiate | Plan Transition, All tasks in Project management process method fragments in OPF, Business Requirements Engineering (from Requirements Engineering process method fragment in OPF) |
| | 2. Create Big Picture | Evaluate Environment Readiness |
| | 3. Create Architecture | Service-Oriented Architecture Engineering |
| | **NA = 3   SMF = 16    MC = 3 / 16 (0.187)** | |
| **Papazoglou** | Task | Corresponding task method fragment(s) |
| | 1. Planning | Plan Transition, All tasks in Project management Process method fragments in OPF |
| | 2. Analysis and Design | Evaluate Environment Readiness, Identify Services, Specify Detail of Services |
| | 3. Construction and Testing | Implement and Test Necessary Services, Implement Necessary Wrappers |
| | 4. Provisioning | Develop Necessary Composite Web Services, Discover Necessary Web Services |
| | 5. Deployment | Publish Web Services |
| | 6. Execution and Monitoring | Monitor Operational Web Services, Compose Web Services Dynamically |
| | **NA = 6   SMF = 16    MC = 6 / 16 (0.375)** | |
| **SDM proposed by Chang and Kim** | Task | Corresponding task method fragment(s) |
| | 1. Identifying business processes | Evaluate Environment Readiness, Business Requirements Engineering (from Requirements Engineering process method fragment in OPF) |
| | 2. Defining Unit services | Identify Services, Specify Detail of Services |
| | 3. Discovering Services | Discover Necessary Web Services |
| | 4. Developing Services | Publish Web Services |
| | 5. Composing Services | Develop Necessary Composite Web Services |
| | **NA = 5   SMF = 16    MC = 5 / 16 (0.312)** | |

## A.1 GAP ANALYSIS

As far as the completeness of the proposed method fragments derived from our study of the eleven prominent service-oriented SDMs is concerned, it should be noted that the proposed set of fragments, as a core for the construction of service-oriented SDMs, may be enhanced further and evolved by the introduction and consideration of any new service-oriented SDMs. The analysis of new service-oriented SDMs can lead to the addition of a new assortment of method fragments too. However, as more and more new SDMs are considered, we expect that the incremental additions to the proposed method fragments diminish marginally. The same argument applies to any other existing service-oriented SDM we had not consider in our research such as the Multi-View SOAD proposed by Kenzi *et al.* [100]. We only claim and showed that the proposed method fragments are complete with respect to the eleven selected prominent service-oriented SDMs.

## B. USABILITY

Having shown the *completeness* of the proposed method fragments in Section VII-A, we must now show that the proposed fragments are *usable* in the construction of situational SDMs based on situational factors of the project at hand. These two properties together validate our proposed method fragments.

A real empirical assessment is required to justify the usability property of the proposed fragments. However, we have two reservations. Firstly, "Software Process Assessment" is still considered as a challenging task in the SME literature [19, 39] and few real case studies can be found to indicate industrial usages [87]. Secondly, performing a wide-range of empirical experiments on the usability of the proposed service-oriented method fragments in several industrial projects and in different software development organizational circumstances would seem to be an ideal way to evaluate our work. However, adopting such an evaluation technique requires considerable amount of time, effort, and recourses to continuously monitor, gather, and measure data during SDM construction. This is not feasible given the time constraint of our research and the unavailability of real projects. Consequently, we expect that the real validity of our proposed fragments should be appraised in the long run. However, our earlier research in this area [67] suggested strongly that original service-oriented SDMs that have been utilized for identifying method fragments have already attested the suitability and applicability of tasks, or better say method fragments because they had been derived from recurrent pre-examined best practices. Therefore, we can assume that our proposed method fragments have been validated too implicitly.

However, to provide a more concrete measurement and explicit evidence on the usability of the proposed method fragments, we conducted two case studies. By usability in the context of SME we mean how much do method fragments satisfy the requirements of an SDM [102]. We define a simple intuitive metric wherein the satisfaction of the requirements of an SDM is defined as the percentage of the number of requirements that are met by method fragments divided by the number of all requirements as shown in Equation III.

$$Usability(\%) = \frac{M}{R} \times 100 \quad \text{(III)}$$

In Equation III, *M* represents the number of requirements met, *R* represents the total number of SDM's requirements, and ***Usability*** represents the percentage of the usability of method fragments. High *Usability* means that method fragments have met most of SDM's requirements. The 100% *Usability* means that method fragments have met all of SDM's requirements. We can thus measure the usability of the proposed method fragments in each case study (i.e., real project) using this criterion.

The two case studies presented here demonstrate how the proposed method fragments were used in the construction of a specific service-oriented SDM based on the enhanced OPF repository. In both case studies, a method engineer first elicited SDM requirements and then designed a SDM by selecting relevant method fragments from the repository. It is of course in this case studies method fragments selection was concerned rather all aspect of SDM construction.

# FIRST CASE STUDY

## A. Scenario

The first case study is the development of a service-oriented system for providing some residential services to employees of an NGO [88]. The NGO has offices in 30 provinces with a total number of 14000 employees. Based on the business process viewpoint, the system should provide online services for booking rooms and accepting payments for the expenses. After deploying the system, any employee can send his/her request to book a room in one of the hotels located in a specific province and track his/her request and pay the expenses by online services provided by third party payment services. Having received the requests, the priorities are automatically determined by the system and a room is assigned to the employee. Employee is informed by the system through email and SMS services and confirms the reservation process.

## B. SDM Requirements

The aim is to satisfy the SDM requirements via appropriate method fragments in order to design the required methodology. Efforts aiming at developing any SDM should begin by clearly defining what the situational requirements of such SDMs are. Method engineers are responsible to map the elicited high-level requirements of the project to method fragments. For simplicity, we envisaged a direct mapping between SDM requirements and method fragments [21]. When the SDM requirements were fixed, method engineers clarified SDM requirements as shown in Table 20. Some requirements were imposed by stakeholders. For instance, business processes modeling and improvement were forced due to the explicit request of stakeholders to receive a detailed documentation of their as-is and to-be business processes. Other requirements were relevant to SDM quality such as agility of development process, fast responsiveness to business volatility, flexibility, time, and cost of system development.

## C. Method Fragment Selection

To illustrate how the proposed method fragments can be really incorporated during construction a service-oriented SDM, we confined ourselves to a simple manual process for method fragments selection rather an automatic method fragments selection with ontology flavor [17]. To realize such SDM requirements through method fragments, method engineers started by setting the overall development life cycle at the highest level of abstraction by using the *Business Optimization Phase*, *Initiation Phase*, *Construction Phase*, *Delivery Phase*, *Usage Phase*, *Retirement Phase* method fragments. All other fain-grain method fragments are placed into phase method fragments. After that, method engineers elaborated the SDM using task method fragments. To do this, method engineers took a set of consecutive inference and decisions based on the requirements and their relations with task method fragments. By considering the sections of each task method fragments, method engineers figured out which task(s) match a requirement. Table 21 synopsizes how each requirement has been satisfied through one or more method fragments. According to this table, analyzing each requirement signifies one or more work products that should be produced to realized a target requirement. So, method engineer selects relevant task method fragments to achieve the required work products. It should be noted that all method fragments need not be included in a project-specific SDM due to project requirements.

The existing OPF repository can be used for requirement elicitation, specification and validation. For such tasks, some of the existing general techniques have been adopted which are most commonly used in any situation and so are incorporated in the constructed SDM. Selection of other tasks is based on the SDM requirements. For instance, method engineers select the *Specify Service Level Agreement (SLA)*, *Discover Necessary Web Services*, *Monitor Operational Web Services* tasks to satisfy #R1. For improving existing business processes, OPF contains numerous tasks that help business processes to be partially or fully optimized. These tasks that are placed in the Business Optimization Phase method fragment assist method engineers to explore organization business processes and re-engineer them based on needs (refer to #R2).

While a number of residency systems had been developed independently in the organization and now they became obsolete, the *Evaluate Environment Readiness* task is selected to assess the documents of the legacy systems to see if they have any asset that can be reused (refer to #R3). The task had significant effect on reducing cost and time of development. Moreover, the old residency system's databases contained a large amount of history records about employees that should have been made available to the new system without losing their integrity. In this regard, the *Plan Transition* task was selected (refer to #R4). As the last functional requirement that the custom SDM should be supported, the *Compose Web Service Dynamically* was selected to satisfy #R5. For instance, e-bank services were replaced by other services while the availability of current service provider was reduced. Selection of some method fragments was unavoidable due to the special situation of the project. For instance, the selection of the *Identify Services* and *Specify Details of the Services* tasks were due to defining and exposing residency business processes as services (refer to #R6).

Having determined the overall development process via selection of appropriate method fragments, we had to show how the chosen tasks had to be performed Method engineers concretized each selected task by associating it with a specific supportive technique (Table 20). For example, to define the right services, method engineers associated *Top-down* and *Bottom–up* approaches to the *Identify Services* task.

Table 20. SDM requirements

| Identifier | Name | Explanation |
|---|---|---|
| #R1 | Utilizing External Services | Organization decided to use third party e-bank services to supply chain of business processes. |
| #R2 | Improving Business Process | The improvement of residency business processes was imperative. |
| #R3 | Using Legacy Systems Services | In order to reduce cost and effort of system development, potential legacy functionalities should be reused. In this regard, a number of old Fox Pro resident systems existed irrespective of being out of date. |
| #R4 | Modernizing Legacy Systems | Existing NGO legacy system and related operational databases should be modernized without stopping the current business processes. Traditional databases should be replaced by novel technologies. |
| #R5 | Conforming to Stated Quality of Services | Quality of external Web Services, specifically full availability and rate of discount per transaction are essential requirements. |
| #R6 | Provide Residency as Service | The residency business process should be exposed as a service to external consumers. |
| #R7 | Requirements-Based | Elicited requirements should be considered in the development of services and consequently the target system. A past unsuccessful experience in NGO domain has shown that a miss- |

| | | understanding of requirements has lead to the development of a useless system | | |

For brevity, responsible roles and related artifacts are not shown in Table 21; they should be defined in real situations. The important point to note is that the resulting methodology must be further refined and adapted iteratively by method engineers during the maintenance of the system in accordance with the project situation through iterative process reviews of the development process.

Table 21. Selected tasks versus SDM requirements

| | Requirement | | Mapping Requirements to Relative Method Fragments | |
|---|---|---|---|---|
| **Identifier** | **Analyzing the Requirement** | **Deduced Required Work Products** | Relative Task Method Fragment(s) | Supportive Technique |
| #R1 | Utilizing external services need to look after for the most appropriate Web-Services. Next, a contract with external supplier to remain on acceptable of QoS should be contracted. Web Services should be monitored during the usage to prevent degrading of QoS. | . A list of candidate Web Service should be discovered on the web. . For selected Web Services a consensus between service provider and consumer should be contracted. . Web Service should be observed during the usage. | Specify SLA | Create SLA contract |
| | | | Discover Necessary Web Services | Search Web Services |
| | | | Monitor Operational Web Services | Monitor the QoS of Web Services |
| #R2 | The current business processes should be modeled, analyzed and re-engineered wherever an improvement is urgent. | . Modeling current business models. . Make improvement on the business process | Process Needs Assessment | Existing Techniques [14,15] |
| | | | Process Tailoring | Existing Techniques [14,14] |
| | | | Process Mandating | Existing Techniques [13,15] |
| #R3 | The feasibility and practicality of currently deployed legacy systems should be assessed whether business logic of existing logic can be wrapped with Web Service technologies while data reside on them. | . A list of candidate business logics can be wrapped into Web Services technology. . State of readiness NGO's infrastructure | Evaluate Environment Readiness | Create a Readiness Report |
| #R4 | While the new system has significant impact on through of the NGO so modernization strategies and alternative solutions should be assessed. | Producing an approved strategy or more strategies to migrate to a new service-oriented system. | Plan Transition | Make Transition Plan |
| #R5 | External Web Service that called via NGO system should be monitored continuously. Ones that work improperly and violate from theirs contracts should be replaced with new Web Services. | . Need to monitor procedure for Web Services adopted in system according to contracts. | Compose Web Service dynamically | Reconfigure Composite Web Services |
| | | | Specify SLA | Create SLA Contract |
| #R6 | The goal of the requirement is to | . A list of candidate | Identify | Top-Down |

| | decompose booking and paying business process into set of service in or to achieve integrity and reusability of process. | service that from residency business process. | Services | Bottom-Up |
|---|---|---|---|---|
| | | | Specify Details of Services | Add Specific Details to Services |
| #R7 | Software's requirements should be formally elicited, documented into requirement engineering documents and then validated by all stakeholders. | . A list of identified and prioritized software's requirements and requirements models. | Requirements Identification | Existing Techniques [14,15] |
| | | | Requirements Prototyping | Existing Techniques [15,15] |
| | | | Requirements Specification | Existing Techniques [14,14] |
| | | | Stakeholder Profiling | Existing Techniques [13,15] |
| | | | Technology Analysis | Existing Techniques [14,15] |

We can now empirically validate the usability of the proposed method fragments using Equation III. According to the table 20, column 1, number of SDM's requirements are 7. In addition, as shown in table 21 column1, number of requirements that are satisfied by one or more method fragments is also 7, . Consequently, all (100%) of the requirements had been met by the method fragments;

$$Usability(\%) = 7/7 \quad (100\%)$$

# SECOND CASE STUDY

## D. Scenario

The second case study adopted from [103] is a Driver Assistance System (DAS) that is categorized in the domain of real-time automotive systems that have high potential for SOSD utilization. The case study was in which target system has multitude of sensors assist driver to monitor the car as much as safe when driver is on the road with her/his car. DAS monitors state of the engine time to time for checking level of oil, pressure of the cylinder heads, stabilization of vehicles and etc,. These checking conducted by sensors which are equipped with safety critical embedded programmed. When a potential crash is recognized by DAS, it informs the driver about that. In reality, detected defect triggers execution of workflow composed of Web-Services that orchestrate Web-Services in order to aid driver to get decision.

Then DAS based on the receiving data from GPS system, aids the driver to select most suitable car services such as garage, tow truck and rental in the area before crash takes place really. The drive specify a list of preference such as desire location of the car service, road conditions, traffic conditions, possible minimum cost for repairing the car that may be earned by various car service companies, alternative ways in order to pay and etc,.

Various garage services provide a set of useful services for driving who are traveling on the road. Services such as make appointment for fixing the car, swap of a part. Tow truck companies also, provide relevant services. A driver whose car is crashed can book an order for moving the car to a garage by truck. For instance, when the driver decides to makes an appointment with a garage, the diagnostic data about car's engine are automatically sent along, allowing the garage to identify the spare parts needed to repair the car. Figure 6, show
From architectural point of view, DAS included a safety critical real-time subsystem as core that check of engine repeatedly and a number of external services that can be utilized when a malfunctioning is near to occurrence.

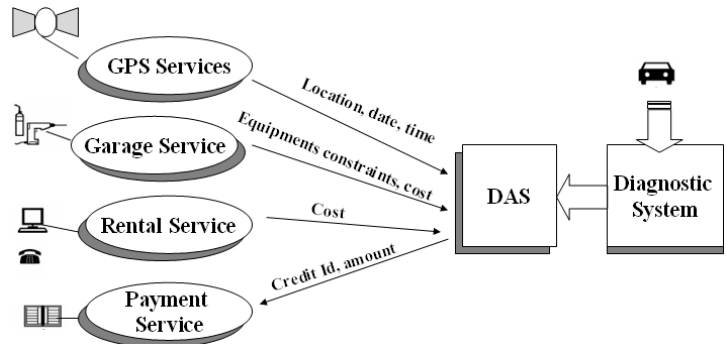

Fig.6. Abstract structure of DAS

## E. SDM Requirements

The policy of software development company about this project is migration from software development (implementation and test) from scratch to assembling approach via utilize existing services. The policy leads company to minimize the time and effort as much as possible.

Some situational factors, led them to this manner. Most of developers in the company are expert in development of data-intensive information systems rather than real time systems. In addition, there is scarcity for obtaining budget of project, developer payment for instance.

From SME point of view, development team needs to define a situational SDM in which a set of consecutive tasks lead them to develop DAS. Method engineer should designate a situational SDM that meet the stakeholders in timely and reasonable manner. He/she also is responsible to discover SDM's requirements and map them to relevant method fragments. Table 22 lists key SDM's requirements that method engineer identified and clarified them.

| | | Table 22. SDM requirements |
|---|---|---|
| Identifier | Name | Explanation |
| #R1 | Constraint on budget | Any code must be implemented, except for trivial parts. Stakeholders adopted policy of utilizing independent and reusable Web-Services that provides more business value for them. |
| #R2 | Incorporating minimum of developers | As sub requirements derived from requirements #R1, stakeholders decided to use a minimum of developers in order to develop the system. |
| #R3 | User's preferences | Driver's preferences should be incorporated for selecting a car service. |
| #R4 | Risk of developer skill | The development team has little developers who are familiar to Web-Service technology. So, they are inclination to be a service consumer. |
| #R5 | Implement and assemble hardware | The low-level code for sensors, timers, analog/digital converter, hardware wrapper and I/O drivers should be designed and |

| | | implemented and tested. In addition, hosted hardware capability should also be tested. |
|---|---|---|
| #R6 | Hardware Validation | Symbolic execution of hardware program also should be performed in order to ensure correctness of code. |

## *F. Method Fragment Selection*

We again confined ourselves to a simple manual process for method fragments selection rather an automatic method fragments selection. Method engineer discerned the dominated approach that is in line with situational factors is to create a composite service constituted a set of fine-grain Web-Services. In this approach the only effort is to find a set of relevant services that intertwined together.

By adopting idea of from high abstraction to lower abstraction, method engineers formed overall development life cycle at the highest level of abstraction by using the *Initiation Phase*, *Construction Phase*, *Delivery Phase*, *Usage Phase* method fragments. All other low level method fragments are placed into phase method fragments.

Afterward, method engineers completed the SDM using task method fragments. For this purpose, method engineers analyzed the sections of each task method fragments and then figured out which task(s) match a requirement. It should be noted #R1, #R2 and #R3 were similar to each other. Consequently, they led method engineer to select same task method fragment. Table 23 synopsizes how each requirement has been satisfied through one or more method fragments.

Table 21. Selected tasks versus SDM requirements

| Requirement | | | Mapping Requirements to Relative Method Fragments | |
|---|---|---|---|---|
| Identifier | Analyzing the Requirement | Deduced Required Work Products | Relative Task Method Fragment(s) | Supportive Technique |
| #R1 | In spite of development embedded code for sensors, other elements of the software systems should be provided via external service. So, obtaining Web-Services from outside reduce cost of project. This leads to searching Web-Services and composing them in order to satisfy user requirements. Developed system is a composite Web-Service that orchestrates a number of fine-grain Web-Services. | A list of candidate Web Service should be discovered on the web based on the driver preferences. | Discover Necessary Web Services | Search Web Services |
| | | SLA of the candidate Web-Service evaluated and those will be selected that satisfy driver requirements. | Specify SLA | Create SLA contract |
| #R2 | This requirement has overlapping with #R1: Utilizing existing exposed Web-Service has significant impact on time and effort of software system development. | A list of candidate web service that meet car driver preference. | Discover Necessary Web Services | Search Web Services |
| #R3 | This requirement is similar to #R1 and #R2. | Similar to R2 | Similar to R2 | Similar to R2 |
| #R4 | This requirement is similar to #R1 and #R2. | Similar to R2 | Similar to R2 | Similar to R2 |

| #R5 | Not supported | - | - | - |
|---|---|---|---|---|
| #R6 | Not supported | - | - | - |

Since, proposed set of method fragments mainly focus on SOSD aspects, method engineer could not find any relevant support for #R5 and #R6. So, #R5 and #R6 were missed for development Driver Assistance. Indeed, this is where OPF repository should be enhanced with specific method fragments for real-time development. According to table 22 column 1, number of SDM's requirements is 6 while number of requirements that were met by the method fragments was 4, table 23, column 1. According to equation III the percentage of requirements satisfaction is:

$$Usability(\%) = 4/6 \quad (66\%)$$

At end of this section, the focus of these case studies were to provide a possible convincing answerer that proposed method fragments are usable to satisfying SDM's requirements. Empirical validation shows that proposed set are usable for construction situaltion SDMs though as with all research, to achive ensure about the research, more projects need to be examained.

## X. CONCLUSION AND FUTURE WORKS

In this research work we presented a set of new service-oriented method fragments that were mined from prominent service-oriented SDMs. These method fragments conform with the OPEN metamodel. We showed how method engineers could select appropriate fragments from the enhanced repository of OPF to effectively construct project-specific service-oriented SDMs.

In this work, we used a number of supportive techniques to derive our proposed method fragments. However, there is a need for more alternative techniques based on the project situation. Moreover, search for new method fragments as an ongoing process is needed. For instance, project management practices in SOSD need a new approach. While a service-oriented software system may be fully developed through a number of distributed development teams, it consequently involves new project management issues in term of team management, cost, and effort estimation. Some future work may thus be stated as (i) enriching the proposed method fragments with more supportive service-oriented techniques, and (ii) searching for other necessary method fragments that are important in new situations and in new software paradigms.